# BESS Optimal Sizing Methodology – Degree of Impact of Several Influencing Factors

Benoît Richard
*Univ Grenoble Alpes,*
CEA, LITEN, DTS, LSEI, INES
F-73370, Le Bourget-du-Lac, France
benoit.richard@cea.fr

Xavier Le Pivert
*Univ Grenoble Alpes,*
CEA, LITEN, DTS, LSEI, INES
F-73370, Le Bourget-du-Lac, France
xavier.lepivert@cea.fr

Yves-Marie Bourien
*Univ Grenoble Alpes,*
CEA, LITEN, DTS, LSEI, INES
F-73370, Le Bourget-du-Lac, France
yves-marie.bourien@cea.fr

*Abstract*—Battery Energy Storage Systems (BESS) are more and more competitive due to their increasing performances and decreasing costs. Although certain battery storage technologies may be mature and reliable from a technological perspective, with further cost reductions expected, the economic concern of battery systems is still a major barrier to be overcome before BESS can be fully utilized as a mainstream storage solution in the energy sector. Since the investment costs for deploying BESS are significant, one of the most crucial issues is to optimally size the battery system to balance the trade-off between using BESS to improve energy system performance and to achieve profitable investment. Determining the optimal BESS size for a specific application is a complex task because it relies on many factors, depending on the application itself, on the technical characteristics of the battery system and on the business model framework. This paper describes a generic simulation-based analytical method which has been developed to determine the BESS optimal size by taking into account both the application and the storage performance over its lifetime. Its implementation and the associated results are presented for two different BESS use cases: A smoothing and peak shaving application for PV injection and an off-grid hybrid microgrid case. In order to provide a better understanding of the most influencing drivers to consider during a BESS sizing procedure, several sensitivity analyses have been carried out on these two illustrative cases. The use of comparative scenarios led to quantify the degree of impact on optimal sizing results of several factors among the following topics: control strategy, forecast quality, degradation of battery performance due to ageing, precision of technical modelling.

## I. INTRODUCTION

Due to the number and variety of services they can provide, energy storage is likely to play a significant role in the optimal mix of flexibility solutions for the European power system. Of the various type of energy storage technology available, battery energy storage systems (BESS) have attracted considerable attention with clear advantages like fast response, controllability, and geographical independence [1,2]. Besides the advantages mentioned, BESS also have a wide scope of applications ranging from short-time power quality enhancement to long-term energy management, as well as reliability enhancement, uninterrupted power supply and transmission upgrade deferral. BESS thus mitigate some of the current and future challenges that grid operators face to improve the overall economics of the infrastructure while reducing the overall carbon footprint and providing reliable services. Specifically, the challenges include managing peak demand, resolving transmission line congestion, and integrating renewable energy technology in a climate of financial risk adversity that will limit new transmission construction.

Over the last decades, significant research and development has been conducted to improve cost and reliability of battery energy storage systems. Although certain battery storage technologies may be mature and reliable from a technological perspective [2], with further cost reductions expected [3], the economic concern of battery systems is still a major barrier to be overcome before BESS can be fully utilized as a mainstream storage solution in the energy sector. The investment costs for deploying a BESS can be significant. That is the reason why, during the implementation of battery energy storage systems, one of the most crucial issues is to optimally determine the size of the battery to define the appropriate balance between the technical improvements brought by the battery and the additional overall cost. Determining the optimal BESS size for a specific application is a complex task because it directly or indirectly depends on a lot of factors, parameters and uncertainties such as:

- Application control strategy, which determines how the storage system is used for the considered application,
- Energy and power application needs over the project lifetime,
- BESS efficiency,
- Degradation of the performances of the battery over its lifetime (capacity and/or power degradation due to battery ageing),
- Uncertainties related to renewable energy sources / forecast errors / load demands / energy prices,
- Realistic economical assessments to correctly evaluate investment as well as operation costs and incomes throughout the project lifetime.

This paper describes a generic method which has been developed to determine the optimal size of a BESS involved in a specific power application. To assess the degree of impact of some influencing factors among those listed above, this method has been used to carry out several sensitivity analyses.

## II. ANALYTICAL METHOD IMPLEMENTED

One of the major objectives of this study is to provide useful synthetic information on the most influencing factors to consider during the BESS sizing procedure. For this purpose, some sensitivity analyses need to be performed to assess how the value of BESS sizing criteria (financial and/or technical metrics) may be affected by a change of influencing parameter, potentially leading to a different optimal size determination. To conduct this survey, an analytical method has been implemented, based on numerical simulation. Although this type of deterministic technique involves significant computational resources to repeat simulations with different combinations of BESS sizes and influencing factors parameters, it provides the necessary flexibility for all criteria and parameter settings required by the sensitivity analysis. Due to deterministic calculation on user-specified intervals, it also enables a better control and visualization of the impact of the factors variation.

### A. Numerical simulation platfom

The method implemented for BESS optimal sizing relies on a numerical simulation platform called SPIDER (Simulation Platform for the Integration of Distributed Energy Resources), which has been developed at CEA in a Matlab / Simulink environment. The main advantages of using this simulation tool both to carry out some techno-economic assessments as required by the BESS optimal size determination and to perform a sensitivity analysis are listed below:

- The Simulink graphic-modelling environment enables to reproduce the functional architecture of the study cases by hierarchical blocks and diagrams and to interface the energy system components models with the control algorithms.

- This platform environment offers a high-level of modularity, enabling to run some similar operation scenarios with different control algorithms and/or different degrees of technical modelling for a same energy component, just by switching the corresponding block model in the Simulink use-case diagram.

- The SPIDER platform benefits from an already existing library of energy component models (PV plants, wind turbines, energy storage systems, battery cells, converters, fuel generators,…) and control algorithm templates developed by CEA.

- The SPIDER platform provides a generic structure and a ready-to-use set of templates, functions and tools for configuring simulations, launching some sensitivity analysis scenarios and customizing post-processing calculations, as well as instantiating the models with some specific set of parameters of the case study.

- The ability to run simulations at different time steps, specified by the user as a configuration parameter.

### B. Simulation-based method for optimal sizing

The general synoptic implemented for the calculation of the optimal sizing indicator of a given configuration (i.e. one particular size of BESS) is depicted on Fig. 1. By repeating this procedure for different BESS configuration sizes, it becomes possible to identify which configuration size leads to the best value of the performance indicator, determining thus the optimal size.

In practice, the optimal sizing tool developed into the SPIDER platform enables to define a range of BESS size values for launching the automatic processing of the defined sizing indicators on the entire search area. At the end of this processing, an overview graphic is produced to visualize the variation of the sizing indicator along with the BESS size, as well as synthetic tables containing all values of intermediate and final indicators for each of the simulated configurations.

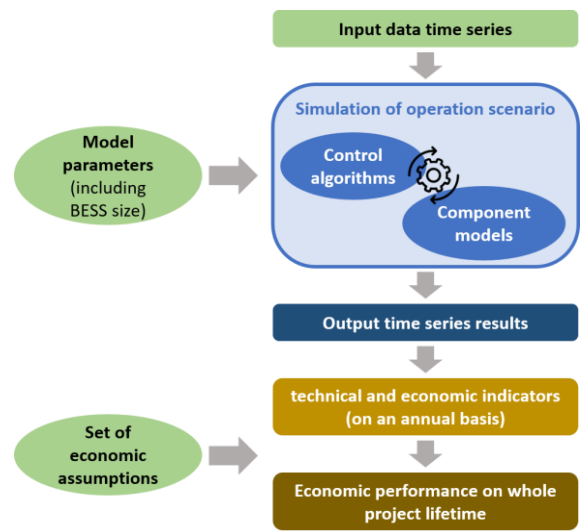

Fig. 1. Overview of the generic method used for BESS optimal sizing

### C. Post-processing calculations

The purpose of post-processing for optimal size determination is to associate a performance value to each simulated configuration, enabling to have a common comparison criteria and identify which configuration, i.e. which size of BESS is the best solution.

Among the different categories of sizing criteria [4], financial indicators are the most commonly used [5,6]. They have the advantage of easily enabling the comparison of different alternatives through a common unit, and can be directly used in discussions with project investors for evaluating financial return and making decisions.

BESS sizing criteria used in the present methodology are based on financial indicators, with the setting of a comprehensive techno-economic assessment to balance the economic value of the rendered service and the total system costs. It relies on the calculation of total system expenses and incomes. Incomes are application dependent since they may come from sales on energy markets, green certificates, feed-in tariffs, etc. For each component, expenses are obtained by summing up investment, total O&M and total replacement costs. O&M costs applied each year and replacement costs applied when equipment estimated life span is expired. The total system expenses are obtained by summing up expenses of all components.

The most appropriate financial indicator for BESS sizing should then be chosen in regards to the purpose of the application. For standalone systems or generation units, LCOE (Levelized Cost of Energy, expressed in €/MWh) is generally well suited as it estimates the average cost of produced energy [6]. However, when the energy application is designed for more complex market rules where variable feed-in tariffs may be applied or ancillary services may be remunerated, other financial indicators like NPV (Net Present Value) or IRR (Internal Rate of Return) may be more suitable [5].

Table I lists the implemented BESS sizing criteria and their definitions. To be able to compute these final key performance indicators according to the formula presented, some annual intermediate indicators need to be calculated from the simulation output time series results, such as the amount of yearly electrical energy generated, annual incomes, total operation costs including replacement costs when necessary, etc.

TABLE I. DEFINITION OF SIZING INDICATORS USED IN THE STUDY

| Performance indicator | Formula | Details |
|---|---|---|
| **Levelized Cost of Energy** <br> It determines the average net present cost of electricity generation over the project lifetime | $$LCOE = \frac{\sum_{n=0}^{N} \frac{CAPEX_n + OPEX_n}{(1+r)^n}}{\sum_{n=0}^{N} \frac{E_n}{(1+r)^n}}$$ | $CAPEX_n$ : total investment costs of year $n$ <br> $OPEX_n$ : total O&M costs of year $n$ <br> $E_n$ : total electrical energy generated in the year $n$ <br> $r$ : discount rate <br> $N$ : project lifetime |
| **Net Present Value** <br> It determines the present value of all future cash flows generated by a project, including the initial capital investment | $$NPV = \sum_{n=0}^{N} \frac{CF_n}{(1+r)^n}$$ | $CF_n$ : cash flow of year $n$ <br> $r$ : discount rate <br> $N$ : project lifetime |
| **Internal Rate of Return** <br> IRR is defined as the discount rate that makes the NPV equal to zero | $$\sum_{n=0}^{N} \frac{CF_n}{(1+IRR)^n} = 0$$ | $CF_n$ : cash flow of year $n$ <br> $IRR$ : internal rate of return <br> $N$ : project lifetime |

Furthermore, this financial evaluation requires a set of economic assumptions that allows realistic estimates of investment, O&M and replacement costs for each component and that include a discount rate value close to the WACC (Weighted Average Cost of Capital) generally observed in the business field of activity of the project. It is worth mentioning that all financial indicators values computed in this study are highly dependent on this set of economic assumptions.

### III. ILLUSTRATIVE APPLICATION USE CASES

The method implemented and the sensitivity study have been carried out for two different illustrative BESS application use cases.

#### A. BESS application #1: PV smoothing and peak shaving

*1) Application description*

This application case corresponds to the call for tenders issued by the French Energy Regulatory Commission in 2015 for installing PV solar plants in French Non-Interconnected Islands. The full description of the call for tenders can be found on the French Energy Regulatory Commission website [7].

For these grid-connected PV power plants, an energy storage system must be installed in order to control the power injection to the grid.

A minimal capacity of the storage system is imposed by the call for tenders, set to 0.5MWh useful capacity per MW of installed PV peak power, at any time of the project life. To respect this constraint, the battery nameplate capacity must be chosen a little higher, to take into account both the deep of discharge (DoD) range and the capacity degradation due to ageing. In our study case, with a DoD of 90% and a capacity degradation up to 30% before battery replacement, the minimal nominal capacity should be set to:

$$rated\ capacity_{min} = \frac{useful\ capacity_{min}}{0.9 * (1 - 0.3)} = \frac{0.5}{0.9 * 0.7}$$

$$\approx 0.8\ MWh\ per\ MW\ of\ PV$$

In order to sell the injected energy at the agreed feed-in tariff, the following main constraints must be respected:

- Power injection profile must be announced in advance: the daily injection profile must be known the day before by the grid operator and the producer shall respect it: outside a tolerance range of ±5% of the installed PV power, some financial penalties are applied.

- PV smoothing: PV fluctuations must be limited to specific ramp rates in the resulting grid injection.

- Power injection during peak period: In order to contribute at the mitigation of the daily peak power demand, the plant must inject energy every day during the two hours of the peak period (19h00 – 21h00) at a minimum power output of 20% of the PV installed capacity. For the energy injected to the grid during the peak period, a bonus equal to 200€/MWh is added to the agreed feed-in tariff.

Fig. 2 provides an illustration of the requested operation for a typical clear day.

*2) BESS optimal sizing*

*a) Sizing criteria*

The economic model of this application, for which the income arises from the sale of electricity at a specified feed-in tariff reduced by any penalties applied when the announced power profile is not respected, leads to choose NPV (Net Present Value) as the sizing indicator.

*b) Optimal sizing reference curve*

Fig. 3 details the system setup used in simulation and Fig. 4 depicts the NPV values obtained along the BESS sizes range explored for the baseline scenario. It shows that the evolution of the sizing indicator as a function of the BESS size has not an optimum curve shape, but is rather quite linear, with the minimum battery size imposed being the most profitable configuration. It can indeed be demonstrated that, because of the economic framework defined by the call of tenders, additional battery capacity costs are always higher than the additional incomes generated by a larger storage system. As the optimal size is always the smallest for this application, this will unfortunately prevent the sensitivity study from identifying when an influencing factor has an impact strong enough to change the value of the optimal size. Nevertheless the impact on the sizing criteria (Net Present Value) could be quantified.

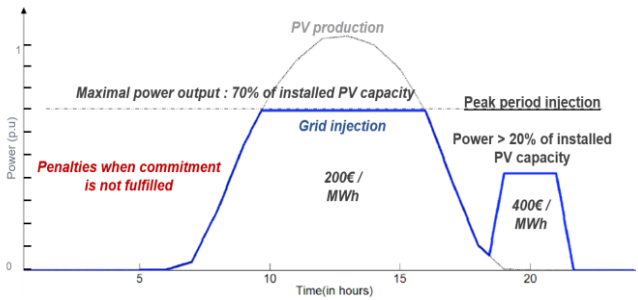

Fig. 2. Application #1 - Daily injection profile (illustration)

| System setup | | |
|---|---|---|
| PV Plant | Installed capacity | 1 MWp (peak power) |
| | PV degradation rate | 0.5% per year |
| | PV producible dataset | 1 year measurement data |
| | PV forecast dataset | 1 year historical irradiance forecasts |
| BESS | Battery technology | Li-ion / app. 6.5 kWh per battery module |
| | Depth of discharge | 90% (from $SOC_{min}$ = 5% to $SOC_{max}$ = 95%) |
| | Battery replacement | Replacement when SOH is 70% |
| | Battery capacity | From app. 800 kWh up to app. 1300 kWh |
| | DC/AC converter power | Up to 700 kVA |

Fig. 3. Application #1 – Simulation setup

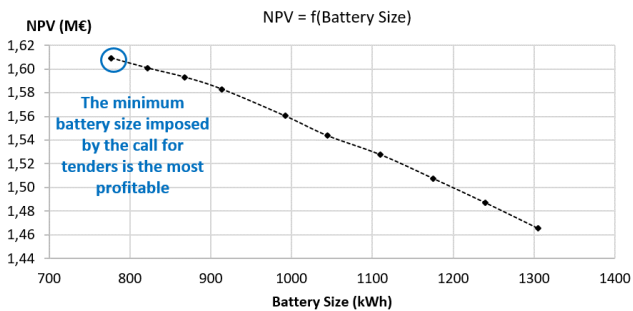

Fig. 4. application #1 - Optimal sizing baseline

### B. BESS application #2: hybrid microgrid

#### 1) Application description

The second application is related to a standalone hybrid microgrid, which is composed of a load to supply for the demand (industrial power profile), and of a diesel generator with the addition of solar PV for the generation, as represented on Fig. 5.

This use case is typical of remote areas where electricity consumption is too low to justify the investment to connect them to the main grid. In the past, fossil fuel generators such as diesel generators have been heavily used for power supply in those areas. However, the rising cost of fuel for the generators, the decreasing cost of renewable energy technologies, as well as the environmental concerns, has led to standalone hybrid energy systems as an attractive solution for remote area power supply.

The key objective of employing a BESS in a standalone hybrid system is to match the imbalance between renewable energy generation and electricity demand to ensure continuity of power supply. In this sense, the functions of diesel generators can be partially or completely replaced by renewable energy and BESS.

#### 2) BESS optimal sizing

##### a) Sizing criteria

For this standalone application case, LCOE is a well suited sizing indicator as it will enable to estimate the average cost of the energy produced to supply the industrial load. The optimal BESS size will be determined by the configuration for which the LCOE is the lowest.

##### b) Optimal sizing reference curve

Fig. 6 details the system setup used in simulation and Fig. 7 depicts the LCOE values obtained along the BESS sizes range explored for the baseline scenario. For this application case, the evolution of LCOE as a function of the BESS size reveals an optimum curve shape. The high LCOE for the smallest BESS configurations [100 – 300 kWh] is mainly composed of OPEX costs, due to the intense use of the fuel generator, and few CAPEX. On the opposite side, the LCOE for the largest BESS configurations [> 700 kWh] contains a strong proportion of CAPEX due to the high initial investment for the storage system procurement, and low fuel costs. Between the two ends of the curve, the optimal trade-off between BESS investment cost and fuel consumption reduction is obtained for a battery size of 440 kWh with the lowest LCOE value, here at 360€/MWh.

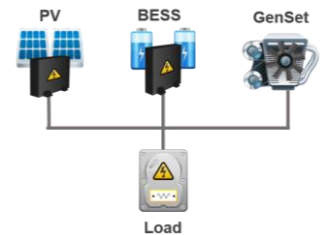

Fig. 5. application #2 - Hybrid microgrid topology

| System setup | | |
|---|---|---|
| Load | Peak power | 40 kW |
| | Load profile | weekly profile from measurement data samples |
| GenSet | Rated power (PRP) | 100 kVA / 80 kWe |
| | Fuel consumption | Consumption @ 110% load (L/h) 25.50<br>Consumption @ 100% load (L/h) 23.50<br>Consumption @ 75% load (L/h) 16.50<br>Consumption @ 50% load (L/h) 11.50 |
| PV Plant | Installed capacity | 130 kWp (peak power) |
| BESS | Battery technology | Li-ion / app. 6.5 kWh per battery module |
| | Battery replacement | Replacement when SOH is 70% |
| | Battery capacity | From app. 110 kWh up to app. 1100 kWh |
| | DC/AC converter power | Up to 200 kVA |

Fig. 6. application #2 – Simulation setup

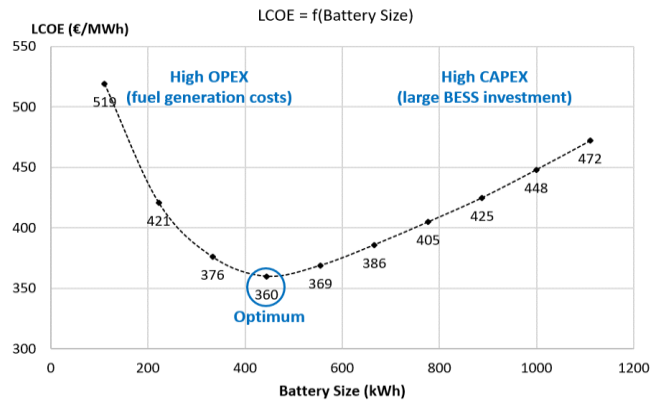

Fig. 7. application #2 - Optimal sizing baseline

## IV. SENSITIVITY ANALYSIS

### A. Sensitivity study scope

To provide some better understanding on the most influencing drivers to consider during a BESS sizing procedure, a sensitivity analysis has been carried out for assessing the impact on BESS optimal sizing of several factors. This study has been performed through the simulation method described in section II, using the two illustrative use-cases described in section III.

To sum up the different areas explored by the sensitivity analysis, Table II lists the different factors which have been investigated with their comparative simulation scenarios.

### B. Illustration of detailed results obtained when assessing the influence of control strategy

In this section is discussed the influence on optimal sizing of the control algorithms which manage the whole energy system in regards to the purpose of the application. For a same application case, different control strategy may be implemented, according to the degree of complexity of the control algorithms which have been developed.

For the hybrid microgrid illustrative case, comparison of sizing results is made by separately processing the simulations with two different sets of control algorithms:

- The basic control strategy has the main function of starting and stopping the diesel generator according to the SOC level of the storage system:
  - When the battery is almost empty (SOC threshold has been set to 10%), the generator is started.
  - When there is a certain quantity of energy in the battery (SOC threshold has been set to 30%), the generator is stopped.

Its secondary function is to regulate the constant balance between the power generation and the electrical consumption of the load. To fulfil this objective, battery capacity is primarily used as a buffer to compensate for any potential imbalances. In the event of excessive generation and a full battery, the PV production may also be reduced.

- The advanced control strategy is more complex as the generator start and stop operations are no longer managed through SOC thresholds but through an optimisation logic which has been implemented thanks to the interface with GAMS software. This optimisation logic aims at minimizing the genset operation costs on a daily horizon. To achieve this objective, the optimisation problem is fed both with a load consumption prediction and a PV forecast, enabling to determine the minimal quantity of additional energy that will be needed from the diesel generator and when it should be produced. As a result, both fuel consumption and genset startup numbers are minimized, leading thus to an operation cost reduction. Concerning the prediction inputs, a real day-ahead forecast has been used for PV and a persistent D+7 prediction has been built for the load consumption as it presents a weekly profile.

As an illustration, Fig. 8 compares the resulting operation for a 3-day simulation period between the two different control strategies: it can be observed that the fuel generator is started less frequently with the advanced control strategy.

Fig. 9 and Fig. 10 show that optimal sizing results are strongly affected by the choice of the control strategy, up to the point of changing the optimal BESS size.

For small BESS configurations where the LCOE is mainly composed of OPEX costs, the reduction in operating costs induced by the advanced control strategy is so substantial that it moves the optimum of the LCOE curve from the BESS size of 440 kWh (LCOE value of 357 € / MWh) to a smallest BESS size of 330 kWh (LCOE value of 344 € / MWh). Consequently, thanks to a higher degree of complexity of control algorithms, not only can a smaller BESS be installed, but also a decrease of the LCOE of 3.64% is achieved.

TABLE II. SENSITIVITY ANALYSIS SCENARIOS

| Influencing factor | Face to face scenarios | |
|---|---|---|
| **Control strategy** | *Baseline* | Basic control algorithms |
| | *Comparative* | Advanced control algorithms (including optimisation) |
| **Forecast quality when predictive control is facing forecast errors** | *Baseline* | PV: standard day-1 forecast<br>Load: persistence day+7 |
| | *Comparative #1* | PV: perfect forecast (actual PV production)<br>Load: perfect forecast (actual consumption) |
| | *Comparative #2* | PV: enhanced forecast with 50% fewer errors<br>Load: enhanced forecast with 50% fewer errors (average between baseline and perfect forecasts) |
| **Precision of the BESS efficiency behaviour** | *Baseline* | BESS model parameters include tables of precise efficiency values varying according to temperature, current and SOC |
| | *Comparative* | BESS efficiency is set up as a constant value (average efficiency) |
| **Degradation of battery capacity due to ageing** | *Baseline* | BESS model parameters include ageing data enabling the simulation to take into account the battery capacity degradation over time |
| | *Comparative* | Battery capacity remains constant over time |
| **Degree of technical modelling of the BESS component** | *Baseline* | In-depth performances battery modelling based on equivalent-circuit equations (EC_model) |
| | *Comparative* | Simplified modelling of the energy/power behaviour of the BESS (E/P_model) |
| **Simulation time-step** | *Baseline* | Time-step of 1 mn |
| | *Comparative #1* | Time-step of 10 mn |
| | *Comparative #2* | Time-step of 1 hour |

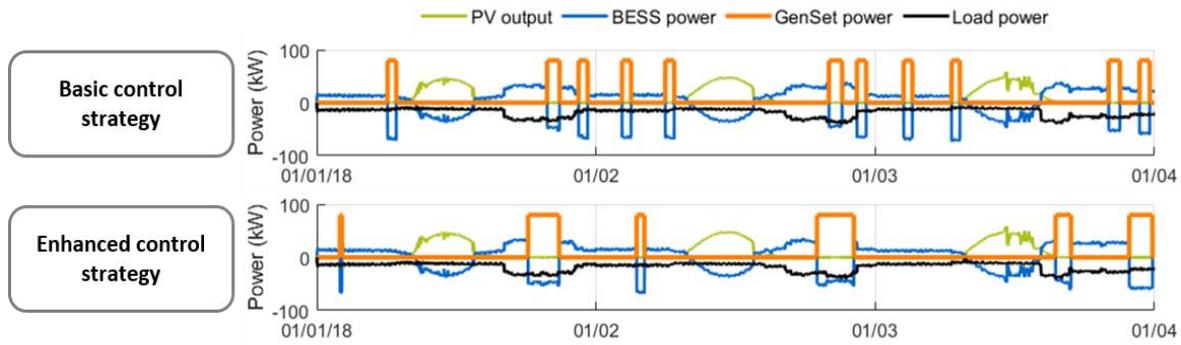

Fig. 8. Comparison of simulated operation with different control strategies

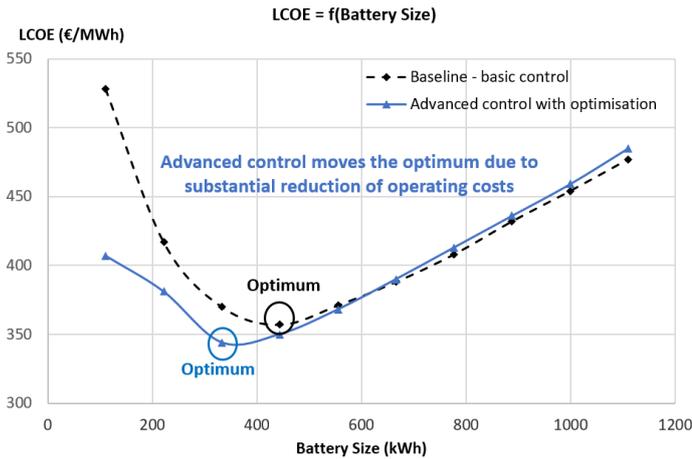

Fig. 9. Influence of control strategy on optimal size determination

Fig. 10. LCOE values obtained through different control strategies

| BESS size (kWh) | Baseline scenario - Basic control strategy LCOE (€/MWh) | Comparative scenario - Advanced control with optimisation LCOE (€/MWh) | LCOE variation |
|---|---|---|---|
| 111 | 528 | 407 | -22,92% |
| 222 | 417 | 381 | -8,63% |
| 333 | 370 | 344 | -7,03% |
| 444 | 357 | 350 | -1,96% |
| 555 | 371 | 368 | -0,81% |
| 666 | 388 | 390 | 0,52% |
| 777 | 408 | 413 | 1,23% |
| 888 | 432 | 436 | 0,93% |
| 999 | 454 | 459 | 1,10% |
| 1110 | 477 | 485 | 1,68% |
| Optimal LCOE variation between the 2 scenarios | | | -3,64% |

For large BESS configurations where the LCOE is mainly composed on CAPEX costs, the enhanced control strategy doesn't bring any advantage since the operating costs are already very low. LCOE is even a little higher: where the basic control can start the diesel generator to charge largest battery capacities (between SOC 10% and SOC 30%) for a period of several days, the optimised control restarts the generator more frequently because of the daily optimisation horizon setup. Some improvements may still be achievable in the advanced control algorithms by fine-tuning the optimisation horizon parameters.

*C. Summary of results obtained for all investigated factors*

For each influencing factor listed in the scope of the sensitivity study, some detailed analyzes similar to the comparison performed in the previous chapter have been carried out for the two different BESS application cases, in accordance with the provided comparative scenarios. The conclusions of the sensitivity study for each factor investigated are summarized in Table III.

V. CONCLUSION

Among the wide range of techniques which can be used to achieve optimal sizing of BESS [4], the present document described a deterministic simulation-based methodology which can be applied for any type of energy system application. Since the main objective of the study was to provide a better understanding of the most influencing factors to consider when determining the optimal size of a BESS, this method was particularly well suited as it offers the adequate level of flexibility to perform various sensitivity analyses.

By using two very different illustrative BESS use cases, the study enabled to:

- illustrate how this generic methodology can be applied to different use cases, for systems composed of various energy components and/or different energy application purposes leading to define different sizing criteria,
- discriminate, among the influencing factors investigated through sensitivity analysis, those whose impact has the same magnitude regardless to the application from those whose impact is application-dependent.

At the stage of modelling or collecting data for optimal sizing purpose, the conclusions provided by this study should help to concentrate the effort on the crucial factors which have the strongest influence on the optimal size determination.

TABLE III. SENSITIVITY ANALYSIS CONCLUSIONS

| Influencing factor | Conclusions |
|---|---|
| Control strategy | **Strong impact**: different control strategies may lead to a different optimal BESS size, as illustrated with the hybrid microgrid application.<br>It is therefore recommended to clearly define the control strategy before determining the optimal size. |
| Forecast quality when predictive control is facing forecast errors | **Highly depends on the application purpose**: if the main function of BESS is to compensate for forecasting errors in the RE sources, as for illustrative application #1, forecast quality is of the highest importance for optimal sizing: a 50% improvement of the forecast quality induced a difference of 15% of the sizing indicator value for application #1. |
| Precision of the BESS efficiency behaviour | **A variable efficiency behaviour can be approximated by an average efficiency single value without any impact on optimal sizing.**<br>However, the average efficiency value must be set precisely since the sizing indicator value is strongly affected by this parameter. An error on BESS efficiency value causes an error bordering on the same magnitude on the sizing indicator. |
| Degradation of battery capacity due to ageing | **Ageing must be taken into account in optimal sizing.**<br>In case of limited availability to precise ageing parameters, an estimation of average degradation is sufficient to obtain appropriate confidence levels on sizing indicators. |
| Degree of technical modelling of the BESS component | **Optimal sizing does not require a high degree of technical modelling**: a simplified model of BESS directly handling power and energy quantities from global efficiency parameters is adapted and leads to the same sizing indicator values, within a one percent interval, as an in-depth performances model based on equivalent-circuit equations. |
| Simulation time-step | The influence of the simulation time-step on optimal sizing **strongly depends on the application time constants** related to the events impacting the operation costs or incomes. An hourly time-step should in general not be recommended as it could lead to an important loss of information about these events. When such events are related to PV fluctuation or fuel generator operation, like on the two illustrative cases, a time-step of 10mn is suitable. |

In addition, these sensitivity study results enable to identify how calculation time can be significantly reduced within an acceptable trade-off between the accuracy of the result and the computing time. As an illustration, by putting together into practice the conclusions related to the use of a simplified BESS model, to the setting of a time-step of 10 minutes and to the ageing estimation by extrapolating a single year simulation, computation time is divided by 840 compared to the baseline scenario, with an average error below 2%. Fig. 11 illustrates the resulting optimal sizing curve superimposed to the baseline reference for application #2 and Fig. 12 compares the LCOE values obtained in both cases as well as the calculation time required for each of the BESS configurations. While each configuration required 1h10mn of computing time with the baseline scenario, it only takes 5 seconds when the conclusions of the study are combined together.

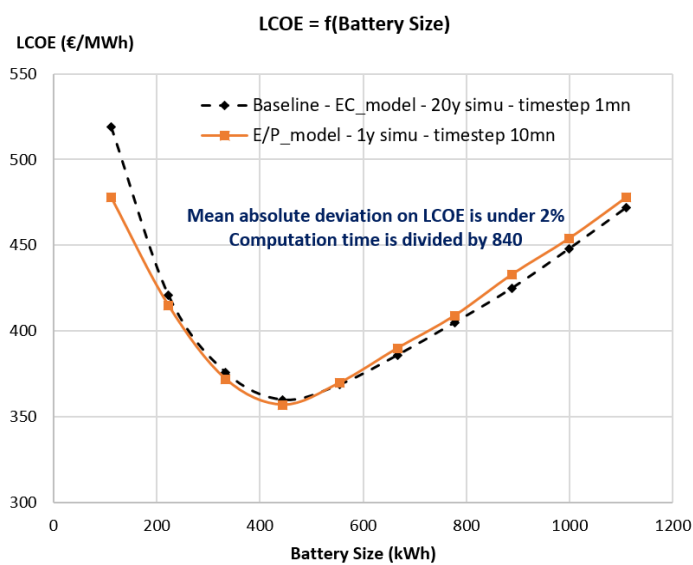

Fig. 11. Conclusive approximation with a time reduction factor of 840

Fig. 12. Detailed results for the conclusive approximation

Despite the fact that computing time can be significantly reduced with relevant approximations, this optimal sizing method, well suited for sensitivity analysis, has certain drawbacks related to its deterministic nature:

- it requires to collect a large amount of data,
- it doesn't take into account the uncertainties of some variables, like production or consumption forecasts,
- a key concern is the need for a large number of simulations with varying battery capacities to reach the optimum solution.

These inherent weaknesses could however be mitigated by further enhancements which could take advantage of the strengths of other techniques such as probabilistic methods or direct search algorithms involving mathematical optimisation or heuristic approach [4].

At last, it should be pointed out that the optimal BESS sizing performed through this study was related to single-function applications. A further way to make the energy capacity (and by extension the physical size of the BESS) a less critical component is the use of advanced dispatch strategies to achieve multiple functions, allowing an existing BESS to be used more effectively and for system design to more effectively use the energy and power capacity of a BESS.

ACKNOWLEDGMENT

The work presented in this paper is part of the results from OSMOSE project, which has received funding from the European Union's Horizon 2020 research and innovation programme under grant agreement No 773406.

REFERENCES

[1] AECOM. (2015). Energy storage study - funding and knowledge sharing priorities.

[2] IRENA. (2015). Battery storage for renewables: market status and technology outlook.

[3] IRENA. (2015). REmap 2030 - Renewable energy prospects: United States of America.

[4] Y. Yang, S. Bremner, C. Menictas and M. Kay, "Battery energy storage system size determination in renewable energy systems: A review" Renewable and Sustainable Energy Reviews, vol. 91, pp. 109 - 125, 2018.

[5] C. Chen, S. Duan, T. Cai, B. Liu and G. Hu, "Optimal allocation and economic analysis of energy storage system in microgrids," IEEE Transactions on Power Electronics, vol. 26, no. 10, pp. 2762-2773, 2011.

[6] C. Shang, D. Srinivasan, T. Reindl, «An improved particle swarm optimisation algorithm applied to batterysizing for stand-alone hybrid power systems» Electrical Power and Energy Systems, vol. 74, pp. 104-117, 2016.

[7] Commission de Régulation de l'Energie (French Energy Regulatory Commission). (2015). 151117CDC_AO_PV_100Plus_ZNI. Retrieved from http://www.cre.fr